\documentclass[twocolumn,preprintnumbers,amsmath,amssymb,aps]{revtex4-1}
\usepackage{graphicx}

\begin{document}

\title{Fermi velocity engineering in graphene by substrate modification}

\author{Choongyu Hwang,$^{1,2}$ David A. Siegel,$^{1,2}$ Sung-Kwan Mo,$^3$ William Regan,$^{1,2}$ Ariel Ismach,$^{4}$ Yuegang Zhang,$^{4}$ Alex Zettl$^{1,2}$}

\author{Alessandra Lanzara$^{1,2}$}
\email[]{ALanzara@lbl.gov}

\affiliation{$^1$Materials Sciences Division, Lawrence Berkeley
National Laboratory, Berkeley, California 94720, USA,}
\affiliation{$^2$Department of Physics, University of California,
Berkeley, California 94720, USA,} \affiliation{$^3$Advanced Light
Source, Lawrence Berkeley National Laboratory, Berkeley, California
94720, USA,} \affiliation{$^4$The Molecular Foundry, Lawrence
Berkeley National Laboratory, Berkeley California 94720, USA.}

\begin{abstract}
The Fermi velocity, $v_{\rm F}$, is one of the key
concepts in the study of a material, as it bears information on a
variety of fundamental properties. Upon increasing demand on the
device applications, graphene is viewed as a prototypical system for
engineering $v_{\rm F}$. Indeed, several efforts have succeeded in
modifying $v_{\rm F}$ by varying charge carrier concentration, {\em
n}. Here we present a powerful but simple new way to engineer
$v_{\rm F}$ while holding {\em n} constant. We find that when the
environment embedding graphene is modified, the $v_{\rm F}$ of
graphene is (i) inversely proportional to its dielectric constant,
reaching $v_{\rm F}\approx$2.5$\times$10$^6$~m/s, the highest value
for graphene on any substrate studied so far and (ii) clearly
distinguished from an ordinary Fermi liquid. The method demonstrated
here provides a new route toward Fermi velocity engineering in a
variety of two-dimensional electron systems including topological
insulators.
\end{abstract}

\maketitle

\section{Introduction}
Due to its lattice structure and position of the Fermi energy, the
low energy electronic excitations of graphene are described by an
effective field theory that is Lorentz invariant~\cite{Kotov}.
Unlike Galilean invariant theories such as Fermi
liquids~\cite{Landau} whose main relevant parameter is the effective
mass, Lorentz invariant theories are characterized by an effective
velocity. Because of this, an increase of electron-electron
interactions induces an increase of the Fermi velocity, $v_{\rm F}$,
in contrast to Fermi liquids, where the opposite trend is
true~\cite{AM}. In the case of graphene, when electron-electron
interactions are weak~\cite{LDA}, $v_{\rm F}$ is expected to be as
low as 0.85$\times$10$^6$~m/s, whereas, for the case of strong
interactions~\cite{GW}, $v_{\rm F}$ is expected to be as high as
1.73$\times$10$^6$~m/s.

Recently, Fermi velocities as high as
$\sim$3$\times$10$^6$~m/s~\cite{Elias} have been achieved in
suspended graphene through a change of the carrier concentration
$n$~\cite{Elias,Aaron,Basov,Andrei}. However, because this
dependence is logarithmic, $n$ needs to be changed by two orders of
magnitude in order to change the velocity by a factor of 3. This
implies that it is unpractical to use $n$ as a way to engineer
$v_{\rm F}$, let alone the fact that one should first realize
suspended graphene in the device~\cite{Elias}. Several other routes
have also been proposed to engineer $v_{\rm F}$ in graphene via the
electron-electron interaction, including modifications of: a)
curvature of the graphene sheet~\cite{Du}; b) periodic
potentials~\cite{CHParNP}; c) dielectric
screening~\cite{DavidPNAS,Fuhrer,Raoux}. While the former two also
substantially modify the starting material, the latter simply
modifies the effective dielectric constant, $\epsilon$, making it
more appealing for device applications~\cite{Antonio}. Despite this
advantage, no systematic study of how to engineer $v_F$ by changing
$\epsilon$ exists to date. Here we provide a new venue to control
the Fermi velocity of graphene using dielectrics, while keeping $n$
constant.

\section{Results}
We perform such a study using three single-layer graphene samples,
which were prepared by chemical vapor deposition (CVD) on Cu,
followed by an {\it in situ} dewetting of Cu on quartz (single
crystal SiO$_2$)~\cite{Ariel} or a transfer onto hexagonal boron
nitride (BN)~\cite{Dean}, and by epitaxial growth on
4$H$-SiC(000$\bar{1}$)~\cite{Hass}. Figures~1{\em A} and~1{\em B}
show angle-resolved photoemission spectroscopy (ARPES) intensity
maps measured near the Brillouin zone (BZ) corner K along the
$\Gamma$-K direction for the two CVD grown samples, which constitute
the first report on Dirac quasiparticle mapping from these samples.
Following the maximum intensity, one can clearly observe almost
linear energy spectra, characteristic of Dirac
electrons~\cite{ShuyunNM}. The momentum distribution curves (MDC),
intensity spectra taken at constant energy as a function of
momentum, are shown in Fig.~1{\em C}. In addition to being
proportional to the imaginary part of the electron self-energy, the
MDC spectral width provides information on the sample quality. A
clear increase of the width is observed by changing the substrate
from SiC(000$\bar{1}$) via BN to quartz, a trend that is in overall
agreement with the theoretical expectation that the electron
self-energy should vary with the inverse square of the dielectric
screening~\cite{Gonzalez_PRL}, as later discussed. The quartz sample
here used constitutes a substantial improvement over a previous
experiment~\cite{exfoliated} on a similar substrate (compare
0.19~\AA$^{-1}$ (red line) versus $\sim$0.7~\AA$^{-1}$ (gray-dashed
line)). The much improved data quality allows for a detailed
self-energy analysis and consequent extraction of important
parameters such as $v_{\rm F}$.

\begin{figure}
  \begin{center}
  \includegraphics[width=1\columnwidth]{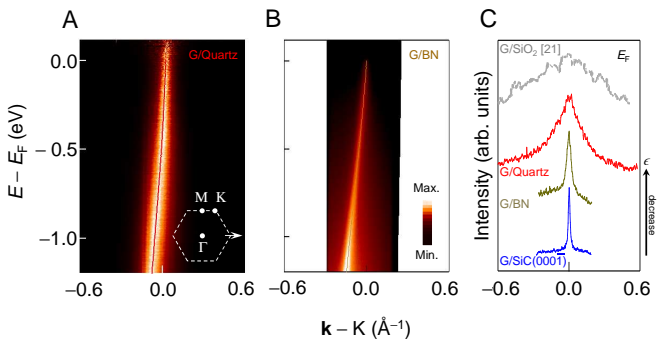}
  \end{center}
\caption{ARPES intensity maps of graphene on quartz and BN. ({\em
A-B}) Normalized and raw ARPES intensity maps of graphene/quartz
(panel ({\em A})) and graphene/BN (panel ({\em B})), respectively.
The red and dark-yellow lines are the dispersions, obtained by
fitting momentum distribution curves (MDCs). ({\em C}) MDCs at
$E_{\rm F}$ for graphene on SiC(000$\bar{1}$) (blue line), BN
(dark-yellow line), quartz (red line), and SiO$_2$~\cite{exfoliated}
(gray-dashed line).} \label{Fig1}
\end{figure}

To understand how the dielectric substrate affects the electronic
properties, in Fig.~2, we show the energy vs. momentum dispersions
for graphene on three different substrates, SiC(000$\bar{1}$), BN,
and quartz, obtained by fitting the MDC spectra. The observed
dispersions exhibit two distinctive features. First, the measured
dispersions deviate from linearity with an increased slope around
$\sim-$0.5 eV for all the samples (compare experimental data to
dashed gray lines in Fig.~2{\em A}). As the substrate is changed
from SiC(000$\bar{1}$) via BN to quartz, corresponding to a decrease
of the dielectric screening, the departure from linearity at high
energy becomes more pronounced. Second, the direct comparison
between experimental dispersions and {\it ab initio} calculations
for the two extreme cases $\epsilon=$~1~\cite{LDA} (suspended
graphene) and $\epsilon=\infty$~\cite{GW} shows another
substrate-dependence (Fig.~2{\em B}). Upon changing the substrate,
the slope increases approaching the dispersion for $\epsilon=$~1.
The deviation from linearity and the enhancement of the slope result
in a reshape of the typical conical dispersion, in a similar fashion
as reported for other charge-neutral graphene
samples~\cite{Elias,DavidPNAS} (see cartoons in the inset of
Fig.~2{\em A}: from left to right). We note that the largest upturn
for graphene/quartz cannot be explained by: a) resolution, which
typically results in the deflection of MDC peaks near $E_{\rm F}$ to
lower momentum, and would involve a much smaller effect by an order
of magnitude ($\leq$a few tens meV)~\cite{resolution_effect}; b) the
presence of other bands with a different azimuthal orientation,
which would cause instead an abrupt increase and a significant
asymmetry of the MDC width at the upturn energy.

\begin{figure}
  \begin{center}
  \includegraphics[width=1\columnwidth]{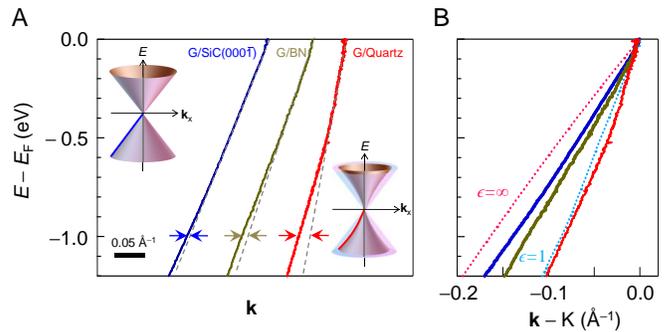}
  \end{center}
\caption{Experimental and theoretical energy spectra for different
dielectric constants. ({\em A}) Experimental dispersions for
graphene on SiC(000$\bar{1}$) (blue line), BN (dark-yellow line),
and quartz (red line). The gray-dashed lines are guides to the eyes.
The insets are cartoons for the electron band structure of graphene
with weak (left) and strong (right) electron-electron interactions.
The data are shifted along the $x$-axis. ({\em B}) The direct
comparison of experimental dispersions with theories:
$\epsilon=\infty$ (magenta line)~\cite{LDA} and $\epsilon=$~1 (cyan
line)~\cite{DavidPNAS}.} \label{Fig2}
\end{figure}

\section{Discussion}
To quantify the effect of dielectric substrates on the
electron-electron interactions and $v_{\rm F}$, we adopt the
standard self-energy analysis to extract self-consistently the
strength of the electron-electron interactions and $\epsilon$
\cite{Kotov,DavidPNAS,Gonzalez,Damascelli}. Figure~3{\em A} shows
the difference between measured dispersions, $E({\rm{\bf k}})$ (from
Fig.~2{\em A}), and the theoretical dispersion for
$\epsilon=\infty$, $E_{\rm LDA}({\rm{\bf k}})$ (shown in Fig.~2{\em
B}). Assuming that electron-electron interactions are effectively
screened for $\epsilon=\infty$, the $E-E_{\rm LDA}$ curve can be
considered a good measurement of the difference between the
self-energy and its value at $E_{\rm F}$. To fit these curves, we
use the marginal Fermi liquid self-energy function as previously
reported~\cite{DavidPNAS,Gonzalez} with an analytic form of
$\frac{\alpha \hbar v_{0}}{4}\,({\rm{\bf k}}-{\rm{\bf k}}_{\rm
F})\,{\rm ln}\frac{{\rm{\bf k}}_c}{{\rm{\bf k}}-{\rm{\bf k}}_{\rm
F}}$ (dotted lines in Fig.~3{\em A}). Here, $\alpha$ is a
dimensionless fine-structure constant (or the strength of
electron-electron interactions) defined as
$\frac{e^2}{4\pi\epsilon\hbar v_{0}}$~\cite{Gonzalez}, $v_0$ the
Fermi velocity for $\epsilon=\infty$,
0.85$\times$10$^{6}$~m/s~\cite{LDA}, ${\rm{\bf k}}_c$ the momentum
cut-off, 1.7~\AA$^{-1}$, and ${\rm{\bf k}}_{\rm F}$ the Fermi wave
number. An overall good agreement with the experimental data is
observed allowing to extract important parameters such as $\epsilon$
and $\alpha$ for graphene on each substrate. For graphene on
SiC(000$\bar{1}$) and BN, we obtain $\epsilon=$~7.26$\pm$0.02
($\alpha=$~0.35) and $\epsilon=$~4.22$\pm$0.01 ($\alpha=$~0.61),
respectively. The extracted value for graphene on BN is in agreement
with the standard approximation $\epsilon=(\epsilon_{\rm
vacuum}+\epsilon_{\rm substrate})/2=$~4.02 and 3.05, where
$\epsilon_{\rm vacuum}=$~1 and $\epsilon_{\rm substrate}=$~7.04 (for
out-of-plane polarization) and 5.09 (for in-plane polarization) in
the low frequency limit (static dielectric constant) for
hexagonal-BN~\cite{BN}. Similarly, the obtained value for graphene
on SiC(000$\bar{1}$) is close to a previous report~\cite{DavidPNAS}.
The apparent discrepancy with the latter (compare
$\epsilon=$~7.26$\pm$0.02 in this work with 6.4$\pm$0.1 in
reference~\cite{DavidPNAS}) is due to the different choice of
reference band (or so-called bare band).  Specifically, in this
work, $E_{\rm LDA}$ is used as the bare band, whereas, in
reference~\cite{DavidPNAS}, the bare band is approximated by a
straight line. Finally, for graphene/quartz, we obtain
$\epsilon=$~1.80$\pm$0.02 ($\alpha=$~1.43), which is smaller than
the expected value of $\epsilon=$~2.45~\cite{SiO}, instead closer to
the experimentally extracted value for suspended graphene
($\sim$2.2)~\cite{Elias}. This observation, together with the
similar energy-momentum dispersion relation at high binding energy
to the theoretical one for suspended graphene (Fig.~2{\em B}), point
to a very weak effect of the substrate.  This is likely a
consequence of the different sample preparation method adopted here
(see Materials and Methods section).

\begin{figure*}[t]
  \begin{center}
  \includegraphics[width=2\columnwidth]{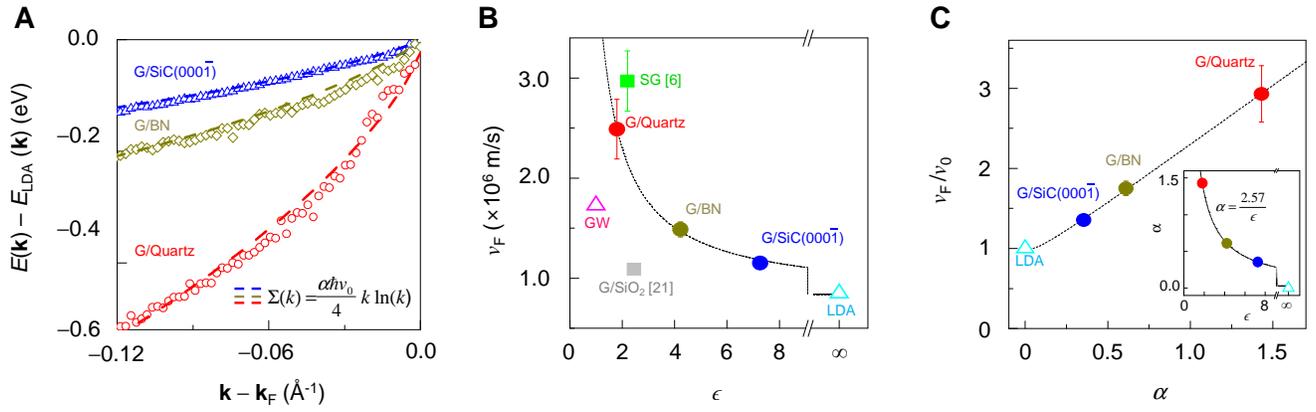}
  \end{center}
\caption{Fermi velocity and the strength of electron-electron
interactions. ({\em A}) $E-E_{\rm LDA}$ dispersions for graphene on
SiC(000$\bar{1}$) (blue line), BN (dark-yellow line), and quartz
(red line). ({\em B}) Fermi velocities as a function of $\epsilon$.
The dashed line is a theoretical curve for $v_{\rm F}$, which is
inversely proportional to $\epsilon$~\cite{Elias,Gonzalez}. Filled
symbols correspond to experimental results, while empty symbols to
theoretical values. $\epsilon=$~2.45 for G/SiO$_2$~\cite{SiO} is
obtained from the standard approximation,
$\epsilon=$~$(\epsilon_{\rm vacuum}+\epsilon_{\rm substrate})/$2
(see text). ({\em C}) The ratio of $v_{\rm F}$, the renormalized
Fermi velocity due to electron-electron interactions, to
$v_0=$~0.85$\times$10$^{6}$~m/s, the bare Fermi velocity in the LDA
limit where $\epsilon=\infty$~\cite{LDA}, as a function of $\alpha$.
The dashed line is the fit given by $v_{\rm F}/v_0=1-3.28\alpha
\{1+(1/4){\rm ln}[(1+4\alpha)/4\alpha]-1.45\}$~\cite{Sarma} for
charge neutral graphene. The inset is the relation between $\alpha$
and $\epsilon$, where the dashed line is a
$\alpha=\frac{e^2}{4\pi\epsilon\hbar v_{0}}=$~2.57/$\epsilon$
curve~\cite{Gonzalez}.} \label{Fig3}
\end{figure*}

In Fig.~3{\em B}, we show the measured $v_{\rm F}$ as a function of
the extracted $\epsilon$ (see also Table~1). Results from a
suspended sample~\cite{Elias} and another graphene/SiO$_2$
sample~\cite{exfoliated} are also plotted for comparison. Upon
decreasing $\epsilon$ from $\infty$ to 7.26 and 4.22, $v_{\rm F}$ is
enhanced from its LDA limit of 0.85$\times$10$^{6}$~m/s (cyan
triangle in Fig.~3{\em B}) to 1.15$\pm$0.02$\times$10$^{6}$~m/s
(blue circle in Fig.~3{\em B}) and 1.49$\pm$0.08$\times$10$^{6}$~m/s
(dark-yellow circle in Fig.~3{\em B}), by 35~\% and 75~\%,
respectively. Surprisingly, when $\epsilon$ is further decreased to
1.80, a dramatic enhancement of $v_{\rm F}$ up to
2.49$\pm$0.30$\times$10$^{6}$~m/s (red circle in Fig.~3{\em B}) is
observed.  Such enhancement corresponds to a 190~\% increase from
its bare value and represents the highest value reported for
graphene on any substrate~\cite{GeimNature,Berger,Sprinkle}.
Interestingly, this velocity is comparable to the value measured for
suspended graphene (green square in Fig.~3{\em B})~\cite{Elias}.
Clearly, a $1/\epsilon$ dependence of $v_{\rm F}$ is observed
(dashed line in Fig.~3{\em B}) in agreement with the theoretical
prediction~\cite{Elias,Gonzalez}. Our result constitutes the first
observation of a power law dependence of the Fermi velocity on the
dielectric constant at fixed $n$. This power law dependence allows
one to achieve, by a smart choice of dielectric, a high value of
$v_{\rm F}$ that cannot be attained otherwise by changing
$n$~\cite{Elias}.

\begin{table}[b]
\caption{Fermi velocity ($v_{\rm F}$), dielectric constant
($\epsilon$), and fine structure constant ($\alpha$) of graphene on
each substrate}
\begin{tabular}{@{\vrule height 10.5pt depth4pt  width0pt}l|ccc}
    \hline
    Substrate & $v_{\rm F}\times$10$^6$~m/s & $\epsilon$ & $\alpha$\\ \hline
    Metals (LDA) & 0.85 & $\infty$ & -\\
    SiC(000$\bar{1}$) & 1.15$\pm$0.02 & 7.26$\pm$0.02 & 0.35\\
    $h$-BN & 1.49$\pm$0.08 & 4.22$\pm$0.01 & 0.61\\
    Quartz & 2.49$\pm$0.30 & 1.80$\pm$0.02 & 1.43\\
    \hline
\end{tabular}
\end{table}

We note that CVD graphene on quartz (red circle in Fig.~3{\em B})
exhibits higher $v_{\rm F}$ than exfoliated graphene on amorphous
SiO$_2$ (gray square in Fig.~3{\em B}) with the same stoichiometry
as quartz. This is a consequence of different sample preparation
process and is due to the larger presence of impurities in the
exfoliated sample, as suggested by the extremely broad spectra (see
gray dashed line in Fig.~1{\em C}). Therefore, although, in theory,
one should expect smaller $v_{\rm F}$ due to screened
electron-electron interactions from impurity~\cite{Fuhrer}, one
should be cautious in extracting meaningful parameters from these
data. We also note that {\it ab initio GW} calculations~\cite{GW}
(magenta triangle in Fig.~3{\em B}) underestimate $v_{\rm F}$ of
suspended graphene. This may be due to the finite $k$-point sampling
inherent in such calculations, or it could also be an indication of
the need to add higher-order terms in the self-energy calculation by
the {\it GW}-approximation.

In Fig.~3{\em C}, we plot the ratio between $v_{\rm F}$ and $v_{0}$,
the expected Fermi velocity in the fully screened case
($\epsilon=\infty$), as a function of $\alpha$. As the strength of
electron-electron interactions is increased, $v_{\rm F}$ is also
enhanced. This is in striking difference with the standard Fermi
liquid picture, where $v_{\rm F}$ is expected to decrease with
increasing $\alpha$~\cite{Polini}. On the other hand, the observed
behavior is consistent with previous theoretical studies for
graphene in the case of specific electron-electron
interactions~\cite{Polini,Sarma} (dashed line in Fig.~3{\em C})
exhibiting the characteristic self-energy spectrum analogous to a
marginal Fermi liquid~\cite{Kotov}. As a result, the departure from
the Fermi liquid picture becomes more important with increasing
electron-electron interactions or decreasing dielectric screening
(see the relation between $\alpha$ and $\epsilon$ in the inset of
Fig.~3{\em C}). Additionally, the observation of $\alpha$ values
close to 1 (neither $\alpha\ll$~1 nor $\alpha\gg$~1) for
graphene/quartz may indicate that a full theoretical treatment
beyond the random-phase approximation~\cite{Kotov} may be required
to understand this sample and/or suspended graphene~\cite{Elias}.

The very good agreement with theoretical
predictions~\cite{Gonzalez,Sarma} for both $v_{\rm F}$ versus
$\epsilon$ (Fig.~3{\em B}) and $v_{\rm F}$ versus $\alpha$
(Fig.~3{\em C}) confirms that the dielectric constants obtained by
the self-energy analysis are self-consistent. Finally the
experimentally determined $\epsilon$ can largely account for the
relatively broad MDCs observed for graphene on quartz (Fig.~1{\em
C}), as compared to graphene on BN and SiC(000$\bar{1}$). For
$\epsilon$ values of 1.80, 4.22, and 7.26, for graphene on quartz,
BN, and SiC(000$\bar{1}$) respectively, the MDC widths, expected to
vary with the inverse square of the dielectric
screening~\cite{Gonzalez_PRL}, should be roughly 16 and 5 times
broader for graphene on quartz and BN than graphene on
SiC(000$\bar{1}$), in line with the experimental observation (see,
for example, Fig.~1{\em C}). We stress that, contrary to a Fermi
liquid system, the broader MDC spectra observed for graphene/quartz
do not necessarily imply decreased transport properties. On the
contrary, the enhanced $\alpha$, the primary cause of the broad
spectra, give rise to an enhancement of Fermi velocity, which is
ultimately one of the most important parameters for device
applications.

In conclusion, we have unveiled the crucial role of dielectric
screening in graphene to control both Fermi velocity and
electron-electron interactions. Additionally, we have shown that
graphene, in its charge neutral state, departs from a standard Fermi
liquid not only in its logarithmic energy spectrum as previously
discussed~\cite{DavidPNAS}, but also in the way that $v_{\rm F}$ is
modulated by the strength of electron-electron interactions. This
dependence provides an alternative way to engineer Fermi velocity
for graphene on a substrate by modifying the dielectric substrate.
This approach can also be applied to charge-doped graphene and other
two-dimensional electron systems such as topological
insulators~\cite{Hsieh} that can be grown or transferred to
dielectric substrates.

\acknowledgments The ARPES measurements and sample growth were
supported by the Director, Office of Science, Office of Basic Energy
Sciences, Materials Sciences and Engineering Division, of the U.S.
Department of Energy under Contract No. DE-AC02-05CH11231.

\section{Materials and Methods}
Graphene samples were prepared in three different ways: epitaxial
growth on the surface of a 4$H$-SiC(000$\bar{1}$) substrate;
chemical vapor deposition (CVD) growth on a Cu film followed by a
transfer onto the surface of boron nitride~\cite{Dean}; and CVD
growth followed by {\em in situ} dewetting of Cu layer in between
graphene and a single crystal SiO$_2$ (namely quartz which is
different from amorphous SiO$_2$ on an Si substrate, the widely used
substrate for exfoliated graphene~\cite{GeimNature})
substrate~\cite{Ariel}. The later procedure is clearly different
from the standard method of exfoliating graphite followed by
deposition onto the amorphous SiO$_2$ layer~\cite{exfoliated}. This
results in a reduced effect of the substrate that is suggested by
the enhanced height variation with respect to the substrate compared
to the sample prepared by the exfoliation and
deposition~\cite{Ariel,Yeh}. The resulting graphene is more
decoupled from the substrate as supported by several features such
as Fermi velocity, dielectric constant, and the electron band at
higher energies closer to suspended sample.

In order to remove any residue including Cu and PMMA, a precursor to
grow CVD graphene and a polymer to transfer graphene, respectively,
we heated the sample to 1000~$^{\circ}$C in ultra-high vacuum. The
removal of Cu is confirmed by: (a) optical microscopy showing a
cleaner image without residual Cu once the sample has been heated;
(b) absence of related Cu features in the ARPES spectra such as 3$d$
electrons at 3.0~eV and 3.5~eV below Fermi energy, and 4$s$
free-electron-like state with a band minimum at 0.25~eV below Fermi
energy~\cite{Patthey}.

High-resolution ARPES experiments have been performed at beamline
10.0.1.1 of the Advanced Light Source at Lawrence Berkeley National
Laboratory using 50~eV photons at 15~K. Energy and angular
(momentum) resolutions were set to be 22~meV and 0.2~$^{\circ}$
($\sim$0.01~\AA$^{-1}$), respectively.


\begin{thebibliography}{34}

\bibitem{Kotov} Kotov V N, Uchoa B, Pereira V M, Castro Neto A H, Guinea F Electron-Electron Interactions in Graphene: Current Status and Perspectives {\it e-print} arXiv:1012.3484v1.
\bibitem{Landau} Landau L (1957) Theory of Fermi-liquids. {\em Soviet Physics JEPT}, 3:920.
\bibitem{AM} Ashcroft N W, Mermin N D (1976) in {\em Solid State Physics} (Saunders College, New York).
\bibitem{LDA} Trevisanutto P E, Giorgetti C, Reining L, Ladisa M, Olevano V (2008) {\em Ab Initio GW} Many-Body Effects in Graphene {\it Phys. Rev. Lett.} 101:226405.
\bibitem{GW} Park C -H, Giustino F, Spataru C D, Cohen M L, Louie S G (2009) Angle-Resolved Photoemission Spectra of Graphene from First-Principles Calculations {\it Nano Lett.} 9:4234-4239.
\bibitem{Elias} Elias D C et al. (2011) Dirac cones reshaped by interaction effects in suspended graphene {\it Nat. Phys.} 7:701-704.
\bibitem{Aaron} Bostwick A et al. (2007) Renormalization of graphene bands by many-body interactions {\it Solid State Commun.} 143:63-71.
\bibitem{Basov} Li Z Q et al. (2008) Dirac charge dynamics in graphene by infrared spectroscopy {\it Nat. Phys.} 4:532-535.
\bibitem{Andrei} Li G, Luican A, Andrei E Y (2009) Scanning Tunneling Spectroscopy of Graphene on Graphite {\it Phys. Rev. Lett.} 102:176804.
\bibitem{Du} Du X, Skachko I, Barker A, Andrei E Y (2008) Approaching ballistic transport in suspended graphene {\it Nat. Nano.} 3:491-495.
\bibitem{CHParNP} Park C -H, Yang L, Son Y -W, Cohen M L, Louie S G (2008) Anisotropic behaviors of massless Dirac fermions in graphene under periodic potentials {\it Nat. Phys.} 4:213-217.
\bibitem{DavidPNAS} Siegel D A et al. (2011)Many-body interactions in quasi-freestanding graphene {\it Proc. Natl. Acad. Sci. USA} 108:11365-11369.
\bibitem{Fuhrer} Jang C et al. (2008) Tuning the Effective Fine Structure Constant in Graphene: Opposing Effects of Dielectric Screening on Short- and Long-Range Potential Scattering {\it Phys. Rev. Lett.} 101:146805.
\bibitem{Raoux} Raoux A et al. (2010) Velocity-modulation control of electron-wave propagation in graphene {\it Phys. Rev. B} 81:073407.
\bibitem{Antonio} Castro Neto A H, Novoselov K (2011) New directions in science ad technology: two-dimensional crystals {\it Rep. Prog. Phys.} 74:082501.
\bibitem{Ariel} Ismach A et al. (2010) Direct Chemical Vapor Deposition of Graphene on Dielectric Surfaces {\it Nano Lett.} 10:1542-1548.
\bibitem{Dean} Dean C R et al. (2010) Boron nitride substrates for high-quality graphene electronics {\em Nat. Nano.} 5:722-726.
\bibitem{Hass} Hass J et al. (2007) Structural properties of the multilayer graphene/4{\it H}-SiC(000$\bar{1}$) system as determined by surface x-ray diffraction {\it Phys. Rev. B} 75:214109.
\bibitem{ShuyunNM} Zhou S Y et al. (2007) Substrate-induced bandgap opening in epitaxial graphene {\it Nat. Mater.} 6:770-775.
\bibitem{Gonzalez_PRL} Gonz\'{a}lez J, Guinea F, Vozmediano M A H (1996) Unconventional Quasiparticle Lifetime in Graphite {\it Phys. Rev. Lett.} 77:3589-3592.
\bibitem{exfoliated} Knox K R et al. (2008) Spectromiscroscopy of single and multilayer graphene supported by a weakly interacting substrate {\it Phys. Rev. B} 78:201408(R).
\bibitem{resolution_effect} Plumb N C et al. (2010) Low-Energy ($<$10 meV) Feature in the Nodal Electron Self-Energy and Strong Temperature Dependence of the Fermi Velocity in Bi$_2$Sr$_2$CaCu$_2$O$_{8+\delta}$ {\it Phys. Rev. Lett.} 105:046402.
\bibitem{Gonzalez} Gonz\'{a}lez J, Guinea F, Vozmediano M A H (1994) Non-Fermi liquid behavior of electrons in the half-filled honeycomb lattice (A renormalization group approach) {\it Nucl. Phys. B} 424:595-618.
\bibitem{Damascelli} Damascelli A, Hussain Z, Shen Z X (2003) Angle-resolved photoemission studies of the cuprate superconductors {\it Rev. Mod. Phys.} 75:473-541.
\bibitem{BN} Geick R, Perry C H, Rupprecht G (1966) Normal Modes in Hexagonal Boron Nitride {\it Phys. Rev.} 146:543-547.
\bibitem{SiO} Gray P R et al. (1984) in {\it Analysis and Design of Analog Integrated Circuits} (Wiley, New York).
\bibitem{GeimNature} Novoselov K S et al. (2005) Two-dimensional gas of massless Dirac fermions in graphene {\it Nature} 438:197-200.
\bibitem{Berger} Berger C et al. (2006) Electronic Confinement and Coherence in Patterned Epitaxial Graphene {\it Science} 312:1191-1196.
\bibitem{Sprinkle} Sprinkle M et al. (2009) First Direct Observation of a Nearly Ideal Graphene Band Structure {\it Phys. Rev. Lett.} 103:226803.
\bibitem{Polini} Polini M, Asgari R, Barlas Y, Pereg-Barnea T, MacDonald A H (2007) Graphene: A pseudochiral Fermi liquid. {\it Sol. State Commun.} 143:58-62.
\bibitem{Sarma} Das Sarma S, Hwang E H, Tse W -K (2007) Many-body interaction effects in doped and undoped graphene: Fermi liquid versus non-Fermi liquid {\it Phys. Rev. B} 75:121406(R).
\bibitem{Hsieh} Hsieh D et al. (2009) Observation of Unconventional Quantum Spin Textures in Topological Insulators {\it Science} 323:919-922.
\bibitem{Yeh} Yeh N -C et al. (2010) Scanning Tunneling Spectroscopic Studies of the Effects of Dielectrics and Metallic Substrates on the Local Electronic Characteristics of Graphene {\it ECS Transactions} 28:115-123.
\bibitem{Patthey} Patthey, F., Schaffner, M. -H., Schneider, W. -D. \& Delley, B. Observation of a Fano Resonance in Photoemission. {\em Phys. Rev. Lett.} {\bf 82}, 2971-2974 (1999).

\end{thebibliography}
\end{document}